\documentclass[twocolumn,letterpaper,aps,prd,superscriptaddress,longbibliography,floatfix]{revtex4-1}

\usepackage{graphicx}	
\usepackage{xspace}	

\begin{document}


\title{Polarization and cross section of midrapidity $J/\psi$ production 
in $p$$+$$p$ collisions at $\sqrt{s}=510$ GeV}

\newcommand{\abilene}{Abilene Christian University, Abilene, Texas 79699, USA}
\newcommand{\augie}{Department of Physics, Augustana University, Sioux Falls, South Dakota 57197, USA}
\newcommand{\banaras}{Department of Physics, Banaras Hindu University, Varanasi 221005, India}
\newcommand{\barc}{Bhabha Atomic Research Centre, Bombay 400 085, India}
\newcommand{\baruch}{Baruch College, City University of New York, New York, New York, 10010 USA}
\newcommand{\bnlcoll}{Collider-Accelerator Department, Brookhaven National Laboratory, Upton, New York 11973-5000, USA}
\newcommand{\bnlphys}{Physics Department, Brookhaven National Laboratory, Upton, New York 11973-5000, USA}
\newcommand{\caucr}{University of California-Riverside, Riverside, California 92521, USA}
\newcommand{\charlesczech}{Charles University, Ovocn\'{y} trh 5, Praha 1, 116 36, Prague, Czech Republic}
\newcommand{\cns}{Center for Nuclear Study, Graduate School of Science, University of Tokyo, 7-3-1 Hongo, Bunkyo, Tokyo 113-0033, Japan}
\newcommand{\colorado}{University of Colorado, Boulder, Colorado 80309, USA}
\newcommand{\columbia}{Columbia University, New York, New York 10027 and Nevis Laboratories, Irvington, New York 10533, USA}
\newcommand{\czechtech}{Czech Technical University, Zikova 4, 166 36 Prague 6, Czech Republic}
\newcommand{\debrecen}{Debrecen University, H-4010 Debrecen, Egyetem t{\'e}r 1, Hungary}
\newcommand{\elte}{ELTE, E{\"o}tv{\"o}s Lor{\'a}nd University, H-1117 Budapest, P{\'a}zm{\'a}ny P.~s.~1/A, Hungary}
\newcommand{\eszterhazy}{Eszterh\'azy K\'aroly University, K\'aroly R\'obert Campus, H-3200 Gy\"ongy\"os, M\'atrai \'ut 36, Hungary}
\newcommand{\ewha}{Ewha Womans University, Seoul 120-750, Korea}
\newcommand{\famu}{Florida A\&M University, Tallahassee, FL 32307, USA}
\newcommand{\fsu}{Florida State University, Tallahassee, Florida 32306, USA}
\newcommand{\gsu}{Georgia State University, Atlanta, Georgia 30303, USA}
\newcommand{\hanyang}{Hanyang University, Seoul 133-792, Korea}
\newcommand{\hiroshima}{Hiroshima University, Kagamiyama, Higashi-Hiroshima 739-8526, Japan}
\newcommand{\howard}{Department of Physics and Astronomy, Howard University, Washington, DC 20059, USA}
\newcommand{\ihepprot}{IHEP Protvino, State Research Center of Russian Federation, Institute for High Energy Physics, Protvino, 142281, Russia}
\newcommand{\illuiuc}{University of Illinois at Urbana-Champaign, Urbana, Illinois 61801, USA}
\newcommand{\inrras}{Institute for Nuclear Research of the Russian Academy of Sciences, prospekt 60-letiya Oktyabrya 7a, Moscow 117312, Russia}
\newcommand{\instpasczech}{Institute of Physics, Academy of Sciences of the Czech Republic, Na Slovance 2, 182 21 Prague 8, Czech Republic}
\newcommand{\isu}{Iowa State University, Ames, Iowa 50011, USA}
\newcommand{\jaea}{Advanced Science Research Center, Japan Atomic Energy Agency, 2-4 Shirakata Shirane, Tokai-mura, Naka-gun, Ibaraki-ken 319-1195, Japan}
\newcommand{\jeonbuk}{Jeonbuk National University, Jeonju, 54896, Korea}
\newcommand{\jyvaskyla}{Helsinki Institute of Physics and University of Jyv{\"a}skyl{\"a}, P.O.Box 35, FI-40014 Jyv{\"a}skyl{\"a}, Finland}
\newcommand{\kek}{KEK, High Energy Accelerator Research Organization, Tsukuba, Ibaraki 305-0801, Japan}
\newcommand{\korea}{Korea University, Seoul 02841, Korea}
\newcommand{\kurchatov}{National Research Center ``Kurchatov Institute", Moscow, 123098 Russia}
\newcommand{\kyoto}{Kyoto University, Kyoto 606-8502, Japan}
\newcommand{\lahorelums}{Physics Department, Lahore University of Management Sciences, Lahore 54792, Pakistan}
\newcommand{\lawllnl}{Lawrence Livermore National Laboratory, Livermore, California 94550, USA}
\newcommand{\losalamos}{Los Alamos National Laboratory, Los Alamos, New Mexico 87545, USA}
\newcommand{\lund}{Department of Physics, Lund University, Box 118, SE-221 00 Lund, Sweden}
\newcommand{\maryland}{University of Maryland, College Park, Maryland 20742, USA}
\newcommand{\mass}{Department of Physics, University of Massachusetts, Amherst, Massachusetts 01003-9337, USA}
\newcommand{\michigan}{Department of Physics, University of Michigan, Ann Arbor, Michigan 48109-1040, USA}
\newcommand{\muhlenberg}{Muhlenberg College, Allentown, Pennsylvania 18104-5586, USA}
\newcommand{\myongji}{Myongji University, Yongin, Kyonggido 449-728, Korea}
\newcommand{\nara}{Nara Women's University, Kita-uoya Nishi-machi Nara 630-8506, Japan}
\newcommand{\natmephi}{National Research Nuclear University, MEPhI, Moscow Engineering Physics Institute, Moscow, 115409, Russia}
\newcommand{\newmex}{University of New Mexico, Albuquerque, New Mexico 87131, USA}
\newcommand{\nmsu}{New Mexico State University, Las Cruces, New Mexico 88003, USA}
\newcommand{\northcg}{Physics and Astronomy Department, University of North Carolina at Greensboro, Greensboro, North Carolina 27412, USA}
\newcommand{\ohio}{Department of Physics and Astronomy, Ohio University, Athens, Ohio 45701, USA}
\newcommand{\ornl}{Oak Ridge National Laboratory, Oak Ridge, Tennessee 37831, USA}
\newcommand{\orsay}{IPN-Orsay, Univ.~Paris-Sud, CNRS/IN2P3, Universit\'e Paris-Saclay, BP1, F-91406, Orsay, France}
\newcommand{\pnpi}{PNPI, Petersburg Nuclear Physics Institute, Gatchina, Leningrad region, 188300, Russia}
\newcommand{\pusan}{Pusan National University, Pusan 46241, Korea}
\newcommand{\riken}{RIKEN Nishina Center for Accelerator-Based Science, Wako, Saitama 351-0198, Japan}
\newcommand{\rikjrbrc}{RIKEN BNL Research Center, Brookhaven National Laboratory, Upton, New York 11973-5000, USA}
\newcommand{\rikkyo}{Physics Department, Rikkyo University, 3-34-1 Nishi-Ikebukuro, Toshima, Tokyo 171-8501, Japan}
\newcommand{\saispbstu}{Saint Petersburg State Polytechnic University, St.~Petersburg, 195251 Russia}
\newcommand{\seoulnat}{Department of Physics and Astronomy, Seoul National University, Seoul 151-742, Korea}
\newcommand{\stonybrkc}{Chemistry Department, Stony Brook University, SUNY, Stony Brook, New York 11794-3400, USA}
\newcommand{\stonycrkp}{Department of Physics and Astronomy, Stony Brook University, SUNY, Stony Brook, New York 11794-3800, USA}
\newcommand{\tenn}{University of Tennessee, Knoxville, Tennessee 37996, USA}
\newcommand{\titech}{Department of Physics, Tokyo Institute of Technology, Oh-okayama, Meguro, Tokyo 152-8551, Japan}
\newcommand{\tsukuba}{Tomonaga Center for the History of the Universe, University of Tsukuba, Tsukuba, Ibaraki 305, Japan}
\newcommand{\vandy}{Vanderbilt University, Nashville, Tennessee 37235, USA}
\newcommand{\weizmann}{Weizmann Institute, Rehovot 76100, Israel}
\newcommand{\wigner}{Institute for Particle and Nuclear Physics, Wigner Research Centre for Physics, Hungarian Academy of Sciences (Wigner RCP, RMKI) H-1525 Budapest 114, POBox 49, Budapest, Hungary}
\newcommand{\yonsei}{Yonsei University, IPAP, Seoul 120-749, Korea}
\newcommand{\zagreb}{Department of Physics, Faculty of Science, University of Zagreb, Bijeni\v{c}ka c.~32 HR-10002 Zagreb, Croatia}
\affiliation{\abilene}
\affiliation{\augie}
\affiliation{\banaras}
\affiliation{\barc}
\affiliation{\baruch}
\affiliation{\bnlcoll}
\affiliation{\bnlphys}
\affiliation{\caucr}
\affiliation{\charlesczech}
\affiliation{\cns}
\affiliation{\colorado}
\affiliation{\columbia}
\affiliation{\czechtech}
\affiliation{\debrecen}
\affiliation{\elte}
\affiliation{\eszterhazy}
\affiliation{\ewha}
\affiliation{\famu}
\affiliation{\fsu}
\affiliation{\gsu}
\affiliation{\hanyang}
\affiliation{\hiroshima}
\affiliation{\howard}
\affiliation{\ihepprot}
\affiliation{\illuiuc}
\affiliation{\inrras}
\affiliation{\instpasczech}
\affiliation{\isu}
\affiliation{\jaea}
\affiliation{\jeonbuk}
\affiliation{\jyvaskyla}
\affiliation{\kek}
\affiliation{\korea}
\affiliation{\kurchatov}
\affiliation{\kyoto}
\affiliation{\lahorelums}
\affiliation{\lawllnl}
\affiliation{\losalamos}
\affiliation{\lund}
\affiliation{\maryland}
\affiliation{\mass}
\affiliation{\michigan}
\affiliation{\muhlenberg}
\affiliation{\myongji}
\affiliation{\nara}
\affiliation{\natmephi}
\affiliation{\newmex}
\affiliation{\nmsu}
\affiliation{\northcg}
\affiliation{\ohio}
\affiliation{\ornl}
\affiliation{\orsay}
\affiliation{\pnpi}
\affiliation{\pusan}
\affiliation{\riken}
\affiliation{\rikjrbrc}
\affiliation{\rikkyo}
\affiliation{\saispbstu}
\affiliation{\seoulnat}
\affiliation{\stonybrkc}
\affiliation{\stonycrkp}
\affiliation{\tenn}
\affiliation{\titech}
\affiliation{\tsukuba}
\affiliation{\vandy}
\affiliation{\weizmann}
\affiliation{\wigner}
\affiliation{\yonsei}
\affiliation{\zagreb}
\author{U.~Acharya} \affiliation{\gsu} 
\author{A.~Adare} \affiliation{\colorado} 
\author{C.~Aidala} \affiliation{\michigan} 
\author{N.N.~Ajitanand} \altaffiliation{Deceased} \affiliation{\stonybrkc} 
\author{Y.~Akiba} \email[PHENIX Spokesperson: ]{akiba@rcf.rhic.bnl.gov} \affiliation{\riken} \affiliation{\rikjrbrc} 
\author{R.~Akimoto} \affiliation{\cns} 
\author{M.~Alfred} \affiliation{\howard} 
\author{N.~Apadula} \affiliation{\isu} \affiliation{\stonycrkp} 
\author{Y.~Aramaki} \affiliation{\riken} 
\author{H.~Asano} \affiliation{\kyoto} \affiliation{\riken} 
% ...
\collaboration{PHENIX Collaboration} \noaffiliation

\date{\today}


\begin{abstract}

The PHENIX experiment has measured the spin alignment for inclusive 
$J/\psi\rightarrow e^{+}e^{-}$ decays in proton-proton collisions at 
$\sqrt{s}$ = 510~GeV at midrapidity. The angular distributions have been 
measured in three different polarization frames, and the three decay 
angular coefficients have been extracted in a full two-dimensional 
analysis. Previously, PHENIX saw large longitudinal net polarization at 
forward rapidity at the same collision energy. This analysis at 
midrapidity, complementary to the previous PHENIX results, sees no 
sizable polarization in the measured transverse momentum range of 
$0.0<p_T<10.0$~GeV/$c$.  The results are consistent with a previous 
one-dimensional analysis at midrapidity at $\sqrt{s}=200$~GeV. The 
transverse-momentum-dependent cross section for midrapidity $J/\psi$ 
production has additionally been measured, and after comparison to world 
data we find a simple logarithmic dependence of the cross section on 
$\sqrt{s}$.
 
\end{abstract}

\maketitle

\section{Introduction}

Measurements of heavy quark bound states in hadronic collisions provide 
unique tools for testing quantum chromodynamics (QCD). Charmonium, the 
bound state of a charm and anticharm quark, is produced predominantly 
via gluon fusion at Relativistic Heavy Ion Collider (RHIC) kinematics. 
The $J/\psi$ is a colorless neutral meson with spin 1 and decays with 
considerable branching ratio into lepton pairs ($\approx 6\%$ each for 
dielectrons and dimuons). Various theoretical models have been developed 
to describe the $J/\psi$ production cross section and polarization as a 
function of transverse momentum, but none can describe both 
simultaneously. All approaches assume a factorization between the 
production of the heavy-quark pair, $Q\bar{Q}$, and its hadronization 
into a meson. Different approaches differ in the treatment of the 
hadronization. The color-evaporation model that is based on quark-hadron 
duality and the color-singlet model (CSM) that only allows hadronization 
without gluon emissions are the most popular earlier approaches; see 
Refs.~\cite{Brambilla:2010cs}~and~\cite{Lansberg:2019adr} for recent 
reviews of theoretical developments and phenomenological work.

One theoretical approach to $J/\psi$ production is within the rigorous 
framework of nonrelativistic QCD (NRQCD), which is an effective theory 
that describes the dynamics of heavy quark bound states at 
nonrelativistic scales
($\nu=v/c\,{\ll}1$)~\cite{Lepage:1992nrqcd,Bodwin:1995hv,Cho:1996nrqcd1,Cho:1996nrqcd2}. The large heavy-quark mass scale relative to the 
hadronization scale, $m_{Q} \gg \Lambda_{QCD}$, factorizes the $J/\psi$ 
production process into the quark-antiquark ($Q\bar{Q}$) pair production 
at short-distance and subsequent formation of the heavy quark meson at 
long-distance. In the former regime, process-dependent short-distance 
coefficients are calculated perturbatively, and in the latter 
nonperturbative regime, the behavior is encoded in long-distance matrix 
elements (LDMEs). The $Q\bar{Q}$ intermediate states are allowed to have 
quantum numbers different from those of the final-state meson. The 
leading-order relativistic corrections, for instance, put the $Q\bar{Q}$ 
either in the color-singlet (CS) state $^{3}S^{[1]}_{1}$ or one of the 
color-octet (CO) states, $^{1}S^{[8]}_{0}$, $^{3}S^{[8]}_{1}$ or 
$^{3}P^{[8]}_{J}$ (J=0,1,2). The relative importance of states with 
different color and angular momentum quantum numbers is estimated by 
$\nu$-scaling rules~\cite{Bodwin:1995hv}. Including CO states is found 
to be crucial as their leading-power contributions $\approx 
\mathcal{O}(\frac{1}{p_T})^4$ show up at leading order (LO), in contrast 
to the CS terms that only appear at next-to-next-to-leading order 
(NNLO). The LDMEs can only be determined from experimental data by 
fitting them via a global analysis. Several groups who performed global 
analyses~\cite{Ma:2011glo,Chao_2012,BUTENSCHOEN:2011g,Gong_2013} 
successfully described world data for transverse momentum ($p_T$) 
spectra, while consistent predictions of the spin alignment measurements 
for quarkonia remain a challenge.

The spin alignment of a positively charged lepton from a $J/\psi$ decay, 
commonly known as ``polarization," has been measured at the 
Tevatron~\cite{Abulencia:2007cdf}, 
RHIC~\cite{Adare:2010tj,Adare:2016jta}, and the Large Hadron 
Collider~\cite{Abelev:2012alice,Chatrchyan:2013cms2,Aaij:2013lhcb,Acharya:2018alice}. 
Measuring spin alignment provides additional tests for the theory and 
understanding dominant quarkonium production mechanisms in different 
kinematic regimes. The $J/\psi$ polarization is measured by fitting the 
angular distribution of a positively charged lepton, shown in 
Eq.~(\ref{eq:angdist}), to data and extracting decay angular 
coefficients.

\begin{equation}\label{eq:angdist}
\frac{dN}{d\Omega} \approx 1+\lambda_{\theta} \cos^{2}{\theta} + \lambda_{\theta\phi}\sin^2{\theta}\cos{2\phi}+\lambda_{\phi}\sin{2\theta}\cos{\phi},  \quad
\end{equation} 

\noindent where the coefficients $\lambda_{\theta}$, 
$\lambda_{\theta\phi}$, and $\lambda_{\phi}$ are determined most 
commonly in the helicity (HX) frame~\cite{JACOB:1959helicity}, 
Collins-Soper~(C-S) frame~\cite{Collins:1977iv} and 
Gottfried-Jackson~(G-J) frame~\cite{Jackson:1964gj} defined in the 
$J/\psi$ production plane. Invariant variables are constructed using 
SO(2) symmetry in choosing the $z$-axis in the production plane. 
Physical interpretation is straightforward only with these invariant 
quantities, making direct comparison possible between experimental 
results in different kinematic regimes.  Equation~(\ref{eq:rotinvvar}) 
shows the two frame-invariant variables defined in 
Refs.~\cite{Faccioli_2010,Faccioli_2011} to characterize the decay 
angular distribution.

\begin{equation}\label{eq:rotinvvar}
\tilde{\lambda} = \frac{\lambda_{\theta} + 3 \lambda_{\phi}}{1- \lambda_{\phi}}, \quad
F = \frac{1 + \lambda_{\theta} + 2 \lambda_{\phi}}{3 + \lambda_{\theta}}
\end{equation}

Recently, general methods have been developed for finding all 
independent invariants under rotations for particles with various spin 
quantum numbers~\cite{Ma:2017rotinv,Martens:2017cvj,Gavrilova:2019rotinv}. 
There are also new theoretical developments for less inclusive observables 
that include looking at the polarization of quarkonia produced in 
jets~\cite{Baumgart:2014upa,Kang:2017yde,Dai:2017cjq,Bain:2017wvk}.  \\

Previously a PHENIX $J/\psi$ measurement at $\sqrt{s}$ = 510~GeV in $p$$+$$p$ 
at forward rapidity showed largely longitudinal net polarization 
(negative angular coefficients)~\cite{Adare:2016jta}, while a prior 
midrapidity PHENIX measurement at $\sqrt{s}$ = 200~GeV was consistent 
with no strong polarization~\cite{Adare:2010tj}. The present analysis 
for midrapidity $J/\psi$ production in $\sqrt{s}$ = 510~GeV collisions 
is complementary to both previous measurements.

\section{Experimental setup}

In 2013, the PHENIX experiment at RHIC collected data from 
longitudinally polarized $p$$+$$p$ collisions at $\sqrt{s}$ = 510~GeV. An 
integrated luminosity of 136 pb$^{-1}$ was used for J/$\psi$ 
polarization measurements at midrapidity.

The PHENIX detector is described in detail in Ref.~\cite{Adcox:2003zm}. 
The PHENIX central arms comprise the east and west arms and cover a 
pseudorapidity range $|\eta| <0.35$ and an azimuthal coverage of 
$\Delta{\phi} = \frac{\pi}{2}$ for each arm. The PHENIX detector 
elements used in this analysis include the drift chamber (DC), pad 
chambers (PC), ring-imaging \v{C}erenkov (RICH) detector and 
electromagnetic calorimeter (EMCal). Charged particle tracks are 
reconstructed with the DC and PC tracking system. These detectors also 
provide the momentum information of the tracks. The data sample used for 
this analysis is the sum of events triggered with three different energy 
deposit thresholds applied in the EMCal. Triggers used in this study are 
described in more detail in Section~\ref{sec:analysis} and in 
Ref.~\cite{phenix_online}. The RICH was used for electron 
identification. Two sets of 64 quartz-crystal radiators attached to 
photomultipliers at $z$ positions of $\pm$ 144 cm and pseudorapidity 
$3.1<|\eta|<3.9$ were used to trigger hard collision events and to 
evaluate the collision vertex position. These beam-beam-counters (BBCs) 
and zero-degree calorimeters were used together to evaluate and 
compare the luminosities seen by PHENIX. In addition, the silicon vertex 
detector (VTX) was placed in the west arm around the beam pipe at 
nominal radii of 2.6, 5.1, 11.8, 16.7 cm with an acceptance of $|\eta| 
<1$ and $\Delta{\phi} = 0.8\pi$. The total material budget expressed as 
a percentage of a radiation length is 13.42\%. As the VTX was not in 
operation in 2013, this created a large source of electron background 
from conversions of direct and decay photons.

\section{Analysis procedure}\label{sec:analysis}
\subsection{Event Selection}

Events were triggered by a minimum energy in any 2$\times$ 2-tower group 
of the EMCal and associated hits in the RICH, in coincidence with the 
minimum-bias trigger condition.  The various EMCal-RICH trigger (ERT) 
thresholds for energy deposited in the EMCal were 2.2, 3.7, and 4.7 GeV. 
Detailed description of the PHENIX ERT system can be found in 
Ref.~\cite{Adare:2010tj}. Different scale-down factors were used for 
each threshold, that is, different fractions of triggered events were 
written to tape. An OR of all these triggers was used for polarization 
measurements and only the lowest threshold trigger was required for 
cross section measurements. The $J/\psi$ candidates are triggered by 
either or both members of the electron-positron pair. The transverse 
momentum of electrons is determined by the bending of the track in the 
magnetic field before the DC. For electron identification, information 
from both the RICH and the EMCal was used.  At least one photomultiplier 
tube in the RICH, associated with the track, is required to have fired. 
The energy $E$ deposited in the EMCal was used for electron 
identification by requiring an expected $(E-p)/p$ ratio within 3 sigma, 
where the measured uncertainty is parameterized as a function of $E/p$. 
The electron identification efficiency is approximately 90--95\%. 
Hadrons misidentified as electrons and conversion electrons are 
subtracted statistically as combinatorial background.  Due to the 
presence of nonoperational silicon vertex detector material located very 
close to the beam pipe, conversion electrons mimic prompt electrons and 
thus direct tagging of conversion electrons would have a very limited 
efficiency; therefore, tagging is not used.

Due to the partial azimuthal coverage on each side of the PHENIX Central 
Arm detector, different arm combinations in the measurements of decay 
pair products give access to different pair $p_T$ regions. For instance, 
high-$p_T$ enhancement is achieved in the sample of $J/\psi$ mesons 
whose decay dielectrons are detected in the same arm due to their small 
opening angle in the lab frame. The dielectrons detected in the east arm 
offer a relatively clean high-$p_T$-enhanced $J/\psi$ sample. The pairs 
within the west arm were not used in this analysis due to excessive 
background consisting of $\gamma$ to $e^+e^-$ conversion caused by 
presence of the VTX that was not operational. The pairs with one lepton 
in each arm were used to obtain $J/\psi$ mesons with lower $p_T$. The 
number of $J/\psi$ events is determined by fitting the invariant mass of 
lepton pairs in the data and counting within the mass range between 2.8 
and 3.3~GeV/$c^2$. Combinatorial background attributed to decay 
electrons from hadrons is eliminated statistically before performing the 
fit using the like-sign method~\cite{Adler:2003qs} established in 
PHENIX. To describe the combinatorial background in the data, the ratio 
of like-sign mass distributions between the data in the same event and 
the mixed-event data is used in normalizing the like-sign mass 
distribution in the mixed-event data. Unlike-sign pairs cannot be used 
for normalization because their mass distributions contain correlated 
background, as described below.  For this reason, the like-sign method 
is also referred to as the mixed event method.

\begin{figure}[!htb]
  \includegraphics[width=1.0\linewidth]{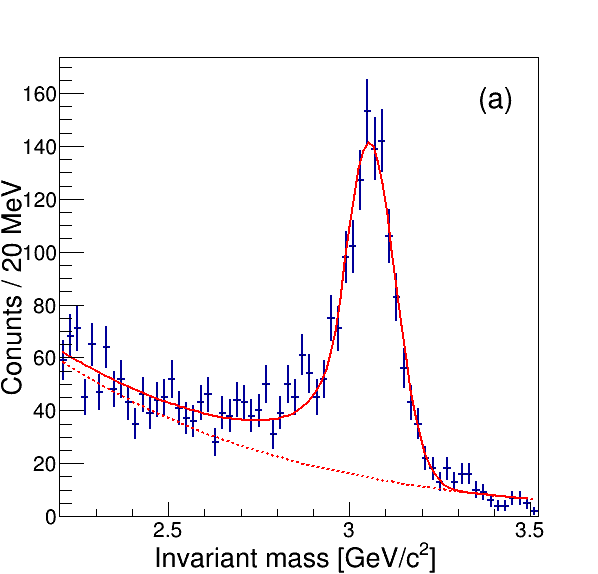}
  \includegraphics[width=1.0\linewidth]{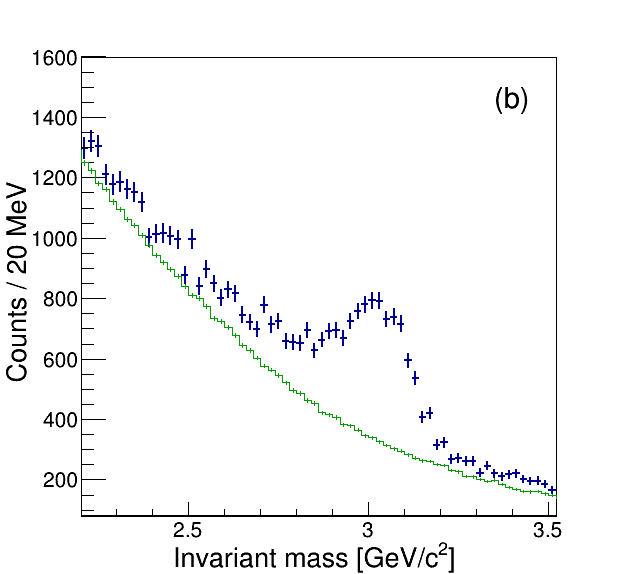}
 \caption{\label{fig:invmass}
(a) Example of invariant mass distribution for the high $p_T$ (3.0 $\leq 
p_T <$ 10.0~GeV/$c$) $J/\psi$ sample. Solid red line indicates a total 
fit and dashed line shows correlated background from the fit. (b) 
Example of invariant mass distribution for the low-$p_T$ (0.0 $< p_T <$ 
3.0~GeV/$c$) sample. Green histogram shows combinatorial background 
calculated using the mixed event method and subtracted before fitting.
}
\end{figure}

The sum of the signal and residual background components, described by a 
crystal ball function~\cite{CBF} and an exponential function, 
respectively, is used to further subtract correlated electron background 
coming from open-heavy-flavor and Drell-Yan decay processes. Invariant 
mass distributions for dielectrons measured in the same arm and opposite 
arms are shown in Fig.~\ref{fig:invmass} along with fit results.

\subsection{Efficiency and acceptance corrections}\label{effacccorr}

The MC pseudodata were generated to estimate the effects of detector 
acceptance and efficiency. Events containing a single $J/\psi$ per event 
were generated with zero polarization, a flat $p_{T}$ distribution, and 
a flat $z$-vertex distribution along the beam direction, using Monte 
Carlo techniques. The $J/\psi$'s subsequently were forced to decay into 
dielectrons within the detector acceptance. The yields were then 
reweighted, as described in more detail below, to emulate effects that 
are present in the data such as smeared vertex distribution along the 
beam axis, the shape of the $J/\psi$ $p_T$ distribution and the ERT 
trigger efficiencies.

\begin{figure*}[!thb]
 \includegraphics[width=0.49\linewidth]{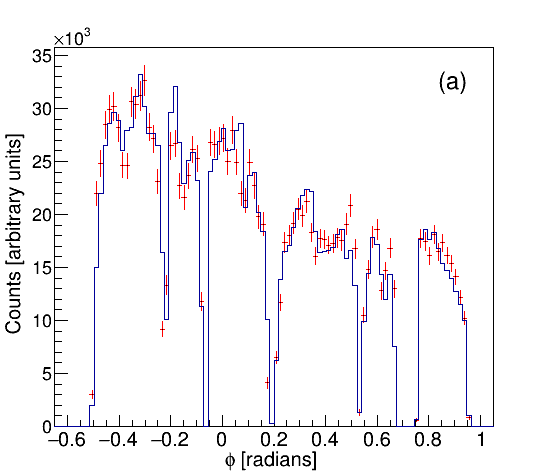}
  \includegraphics[width=0.49\linewidth]{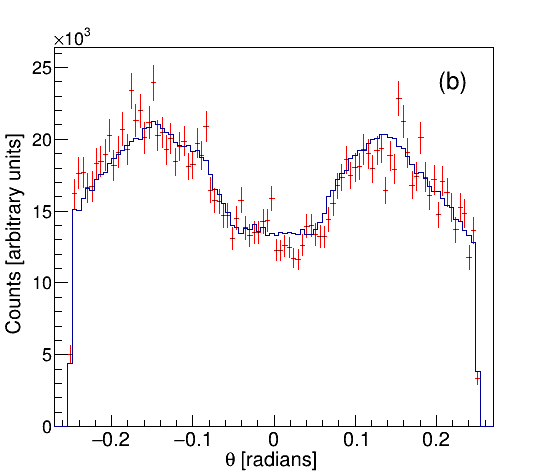}
 \caption{\label{fig:simdatacomp}
Comparison of azimuthal angle $\phi$ (a) and polar angle $\theta$ (b) 
distributions in data (red points) and MC (blue histogram) for 
electrons.
 }
 \end{figure*}

Polarization measurements are particularly susceptible to inconsistency 
between the simulation and data as the acceptance corrections accounting 
for the zero polarization baseline are solely dependent on the 
simulation. A comparison of the polar and azimuthal angular 
distributions between the data and MC simulation, after the effects of 
dead areas, shows good agreement, as seen in Fig.~\ref{fig:simdatacomp}.

 \begin{figure}[!th]
 \includegraphics[width=1.0\linewidth]{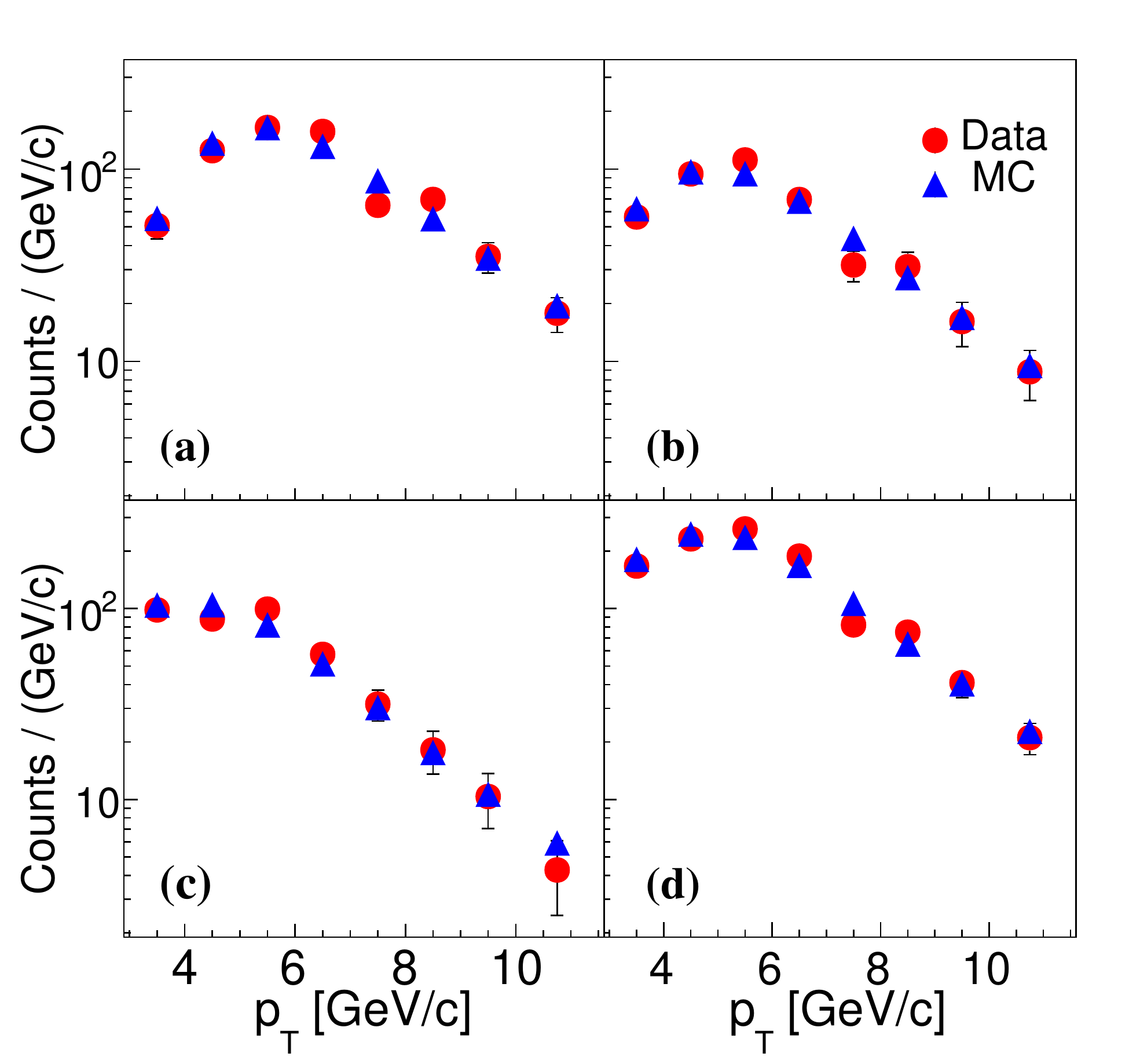}  
 \caption{\label{fig:pttrigs}
Comparison of $J/\psi$ $p_T$ shape in the data with MC for events 
triggered with three different ERT settings [(a) Type A, (b) Type B, (c) 
Type C] and (d) for all settings inclusive Type A+B+C. For direct 
comparisons, trigger efficiency corrections are applied to MC to 
describe the data.
 }
 \end{figure}
 
The shape of $\frac{dN (J/\psi)}{dp_T} $ can affect the polarization 
extraction in a nontrivial manner due to the limited azimuthal 
acceptance of the PHENIX Central Arms. Opposite-sign pairs bending 
toward each other into the detector populate different phase space than 
the ones bending away from each other. For this reason, the $p_T$ 
differential $J/\psi$ cross section has been measured in data. The 
measured cross section was fit with a Kaplan function, defined in 
Eq.~\ref{eq:kaplan}, and the fit results were then used to reweight the 
MC pseudodata for the acceptance correction.

\begin{equation}\label{eq:kaplan}
\frac{d\sigma}{dp_T} = \frac{A \cdot p_T}{\left[1+\left(p_T/b\right)^2\right]^{n}}
\end{equation} 

As the data were sampled from an OR of three triggers with different 
energy thresholds, a method was developed to emulate the trigger 
efficiency effect on the $p_T$ shape of the data. The efficiency of 
triggering on $J/\psi$ events for each type of ERT was determined from 
the minimum-bias dataset by checking the trigger condition. The 
acceptance-corrected pseudodata were then processed by randomly sampling 
the distributions of ERT-measured efficiencies and prescale factors. The 
detector area masked due to being ineffective or suffering from 
excessive electron background was taken into account in this method. 

The shape of the raw $J/\psi$ yield as a function of $p_T$ in data 
before corrections is in excellent agreement with the efficiency 
corrected normalized yield in simulation for each trigger type as well 
as for all possible trigger combinations (see Fig.~\ref{fig:pttrigs}).
 
\subsection{Cross Sections} 

For cross section measurements, the raw $J/\psi$ yield is corrected for 
trigger efficiencies as well as track reconstruction efficiencies and 
acceptance. The EMCal trigger efficiencies were measured using data and 
the electron reconstruction efficiencies and acceptance corrections were 
obtained from MC simulation as described in the previous section. An 
integrated luminosity of 136 pb$^{-1}$ was sampled in the analysis and 
the global normalization of the cross section was determined by the 
$p$+$p$ cross section sampled by the BBC trigger, which is found to be 
32.5 $\pm$ 3.0 (stat) $\pm$ 1.2 (syst) mb, based on van-der-Meer scan 
results. The effects of multiple collisions per beam crossing due to 
high luminosity at $\sqrt{s}$ = 510~GeV, which were estimated to be of 
order of 20\%, were also taken into account in counting $J/\psi$ yields.

\subsection{Angular coefficients} 

\begin{figure}[thb]
 \includegraphics[width=1.0\linewidth]{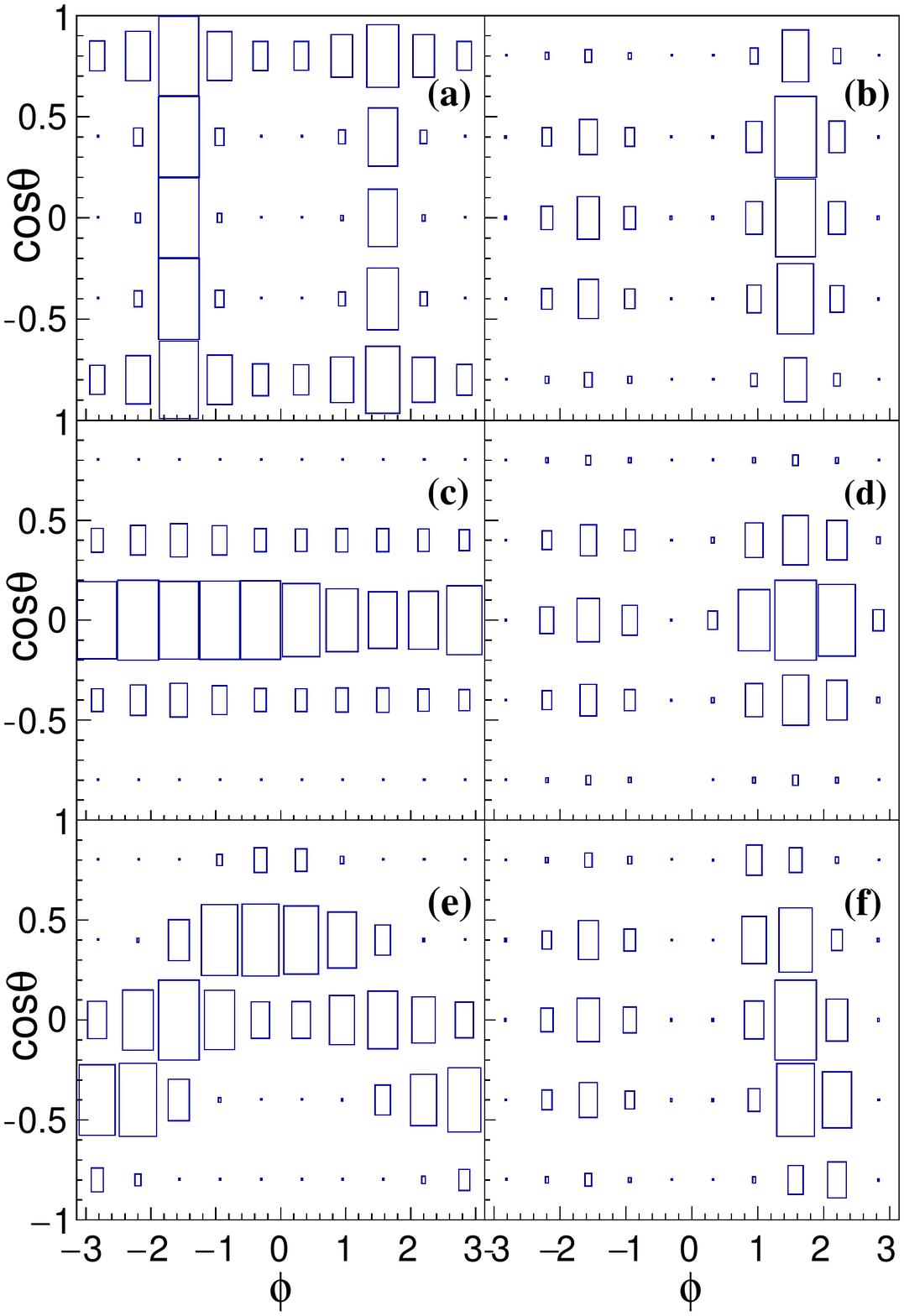}  
 \caption{\label{fig:angdist}
Polar-azimuthal angular distributions in three different polarization 
frames [(a)(b) Helicity, (c)(d) Collins-Soper, (e)(f) Gottfried-Jackson] 
for decay positrons in MC pseudodata generated with a flat $J/\psi$ 
$p_T$ and flat z-vertex distribution.  The distributions are reweighted 
for $p_T$ and z-vertex distribution in the data and corrected for 
trigger efficiencies. The $p_T$ ranges are (a)(c)(e) 
$0.0<p_T<3.0$~GeV/$c$ and (b)(d)(f) $3.0\leq p_T<10.0$~GeV/$c$.
 }
 \end{figure}

To account for acceptance and efficiency effects, the raw 
angular distribution in data is divided by the reweighted single 
$J/\psi$ MC pseudodata generated under an assumption of no polarization 
as described in Section~\ref{effacccorr}. This procedure closely follows 
methods described in Ref.~\cite{Adare:2016jta} and additionally adopts 
the trigger emulator described in Section~\ref{effacccorr}. The 
$\cos{\theta}-\phi$ distributions of the decay positron in three 
different polarization frames in MC pseudodata are shown in 
Fig.~\ref{fig:angdist}.

\begin{figure}[!thb]
 \includegraphics[width=1.0\linewidth]{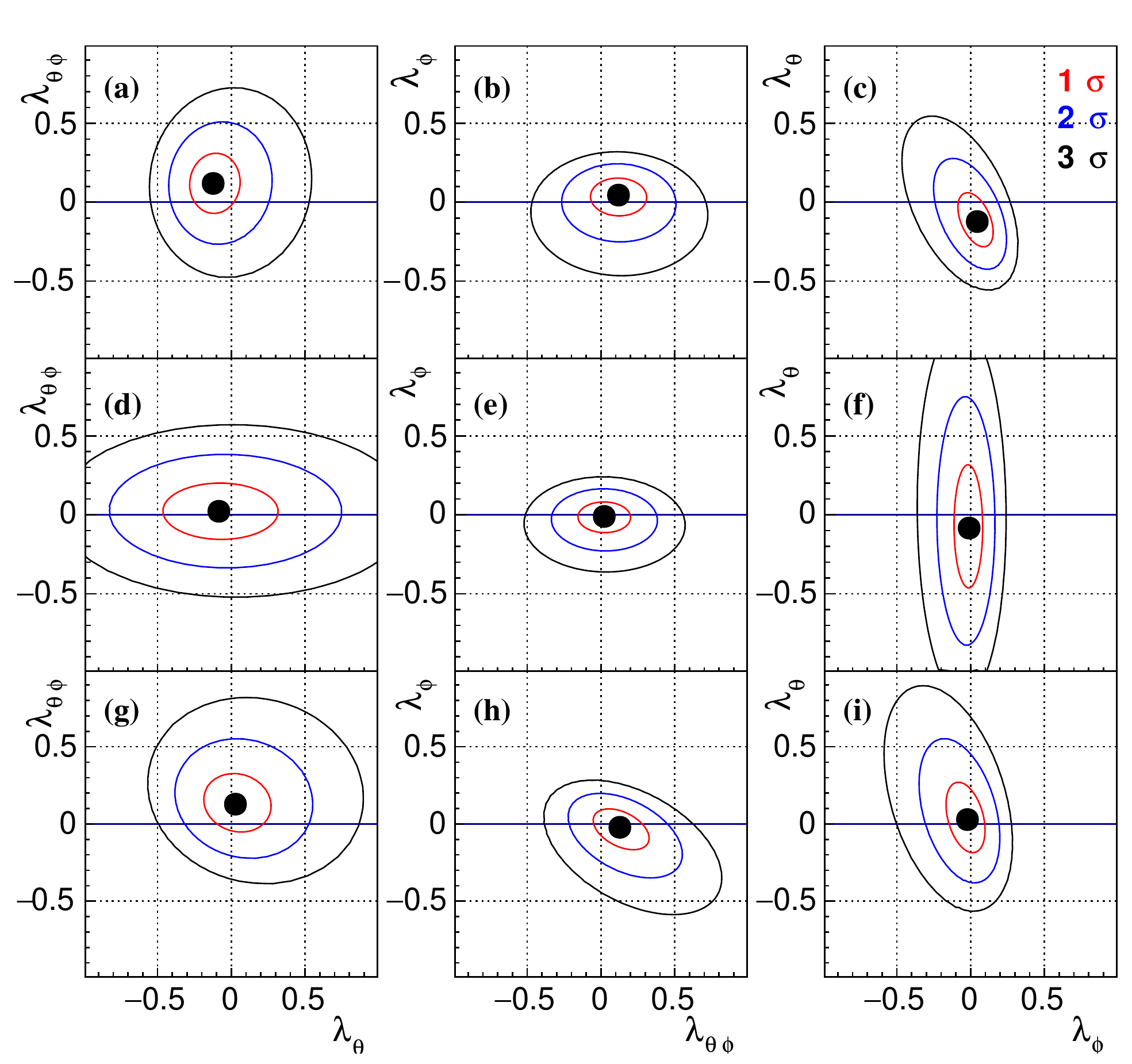}  
 \caption{\label{fig:fitcontours}
The best-fit results for the three angular coefficients are shown with a 
central value in a dot along with contours at 68.3\%, 95.4\% and 99.7\% 
confidence levels. The polarization frames are (a)(b)(c) HX, (d)(e)(f) 
C-S, and (g)(h)(i) G-J. The plot is for the high $p_T$ bin (3.0$\leq 
p_T<$ 10.0~GcV/$c$).
 }
 \end{figure}

Angular coefficients were extracted with three different methods; the 
$\chi^2$ method, the maximum log-likelihood (MLL) method and the 
parametric bootstrap method~\cite{Hastie_01}. The $\chi^2$ fitting 
results are displayed in contours at 68.3\%, 95.4\% and 99.7\% 
confidence level (CL) in Fig.~\ref{fig:fitcontours}. One can see that 
some of the coefficients are correlated in one frame and uncorrelated in 
other frames. The helicity frame is known to be orthogonal to the C-S 
frame and similar to the G-J frame at midrapidity. This is also seen in 
our results, i.e.~the correlation pattern between $\lambda_{\theta}$ and 
$\lambda_{\phi}$, represented by the orientation of elliptic fit 
contours, in the helicity frame appears to be better aligned with the 
ones in the G-J frame. The MLL method treats low statistics bins more 
properly than the $\chi^2$ method, while it is prone to underestimating 
uncertainties as the signal-to-background ratio becomes smaller. In 
fitting methods, the uncertainties on the invariants that are a function 
of primary angular coefficients are propagated from the ones on 
coefficients, taking into account the correlation matrix. The parametric 
bootstrap method properly estimates uncertainties on the invariant 
variables.  In this method, a Gaussian noise term is added to each 
measurement point at each sampling trial with a standard deviation of 
the corresponding measured uncertainty. After a large number of sampling 
procedures, the uncertainties are defined at a 68\% CL.

The results are consistent among different methods within 0.5$\sigma$ in 
all frames and the central measured values and statistical uncertainties 
from the best-fit results were taken as the final results.

At lower $p_T$ ($<$3.0~GeV/$c$), the decay electron and positron are 
detected in opposite PHENIX central arms. A silicon vertex detector 
(VTX) was installed in the PHENIX west arm, but was not operational 
during the data taking. This resulted in a high level of irreducible 
electron background originating from photon conversion in the VTX 
material. These random combinations of background electrons were 
subtracted using an event mixing technique, but resulted in higher 
statistical and additional systematic uncertainty for the low $p_T$ 
measurement. The PHENIX acceptance in $\cos{\theta}-\phi$ space is very 
different for decay electrons and positrons detected in opposite PHENIX 
Central Arms compared to the case when both the electron and positron 
are detected in the same arm. As a result, the acceptance for the C-S 
frame turned out to be very limited, and a polarization measurement in 
this frame was not done at low $p_T$. The polarization coefficients at 
low $p_T$ were determined using only the $\chi^2$ fitting method.

\subsection{Systematic uncertainties}

Sources of systematic uncertainties on the polarization and cross 
section measurement include uncertainties on acceptance, tracking and 
electron identification efficiency, trigger efficiencies, uncertainties 
on the global normalization derived from the cross section seen by the 
BBC, and the trigger rate dependence of $J/\psi$ yield counting. The two 
latter sources do not affect the polarization measurements at all, but 
they are the main sources of systematic uncertainty for $J/\psi$ cross 
section determination.

During RHIC run 2013, the proton beam luminosity was very high, which 
resulted in a nonnegligible probability of multiple collisions per beam 
crossing at high trigger rates. The correction to the $J/\psi$ cross 
section due to multiple collisions per crossing was calculated by 
dividing the whole data set into five run groups with different trigger 
rates, calculating the cross section for each group, and extrapolating 
to zero trigger rate. The correction was estimated to be $0.80 \pm 
0.20$. The large uncertainty of this estimate is due to extrapolating 
from very high trigger rates to zero and lack of statistics. This is the 
largest systematic uncertainty of the $J/\psi$ cross section.

The primary source of systematic uncertainties on the polarization 
measurements in the central-arm detectors is the shape of the $J/\psi$ 
$p_T$ spectra. The $p_T$ shape is affected by statistical uncertainties 
on the raw measured yields as well as uncertainties on the ERT 
efficiencies. The uncertainties on the ERT efficiencies were estimated 
using a sampling method. In each sampling trial, parameters were 
extracted from fitting the efficiency curve for each trigger type, and 
the analysis was repeated with those parameters to estimate the combined 
ERT efficiencies. The quadratic sum of the statistical and systematic 
uncertainties was assigned as a total uncertainty at each measurement 
point. The systematic uncertainties on angular coefficients that are 
attributed to $p_T$ shape were estimated using the parametric bootstrap 
method with a Gaussian noise term corresponding to the aforementioned 
uncertainties. Uncertainties estimated in this method depend on the 
$p_T$ of $J/\psi$, the type of $\lambda$ coefficient and the 
polarization frame. Resulting uncertainties from this source are larger 
(as high as $\approx$0.5 on $\lambda_{\theta}$) at $p_T <$ 3.0~GeV/$c$ and 
$<$0.1 in most cases at $p_T \geq$ 3.0~GeV/$c$. In addition, the 
differences between fitting methods, estimated to be less than 0.5 times 
the statistical uncertainties in all angular coefficients, were added as 
systematic uncertainties at high $p_T$.

At low $p_T$ (below $\approx$3.0 GeV/$c$), the PHENIX acceptance in 
$\cos{\theta} - \phi$ space is very different from that at higher $p_T$. 
Relative contributions from different sources of systematic 
uncertainties also change, while the dependence on exact $p_T$ shape 
remains a dominant factor. An additional systematic uncertainty at low 
$p_T$ is caused by the much larger combinatorial background, which had 
to be subtracted using a mixed-event method. This uncertainty was 
estimated by varying the mixed-event background normalization by 
$\pm$1$\sigma$ and was found smaller than the two uncertainties 
mentioned above. Systematic uncertainties of this measurement are shown 
in Table~\ref{table:polresults}.

\section{Results}\label{result}

\begin{figure}[thb]
 \includegraphics[width=1.0\linewidth]{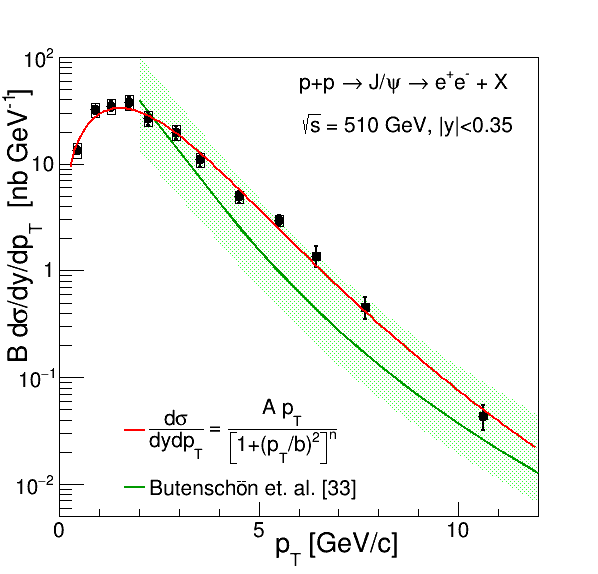}  
 \caption{\label{fig:jpsixsection}
The measured differential $J/\psi$ cross section times branching ratio 
as a function of transverse momentum. Fit parameters are $A = 37.6 \pm 
2.2$~nb/(GeV/$c$), $b = 4.33 \pm 0.28$~GeV/$c$, and $n = 4.61 \pm 0.32$. 
Open rectangles show systematic point-by-point uncertainties. Global 
normalization uncertainty of 10.1\% is not shown. Green curve shows 
theory prediction based on full NRQCD at NLO with leading relativistic 
corrections that include CS and CO states, provided by Butensch\"on et 
al.~\cite{Butenschoen:2011c}. Light green band indicates theory 
uncertainty.
}
 \end{figure}

The measured $p_T$-differential $J/\psi$ cross section is shown in 
Fig.~\ref{fig:jpsixsection} with a fit to a Kaplan function. Due to the 
higher collision energy the $p_T$ spectrum is harder than previously 
published PHENIX results for $\sqrt{s}$ = 200~GeV \cite{Adare:2010tj}, 
where the fit parameter $b$ that determines the hardness of the spectrum 
for a given $n$ was estimated to be smaller at 3.41 $\pm$ 0.21 and the 
parameter $n$ was comparable at 4.6 $\pm$ 0.4. 
Figure~\ref{fig:jpsixsection} also shows a comparison of the measured 
$J/\psi$ differential cross section with a theory prediction based on 
full NRQCD at NLO with leading relativistic corrections that includes CS 
and CO states, provided by Butensch\"on et al.~\cite{Butenschoen:2011c}. 
The sources of theory uncertainties include variations of theory scale 
and LDMEs. Within its uncertainties, the theory calculation is in 
agreement with the experimental results within its valid range of $p_T 
\gtrsim 2$~GeV/$c$, justifying use of the theory model for predictions 
of polarization measurements in this kinematic range.

\begin{figure}[!tbh]
\includegraphics[width=1.0\linewidth]{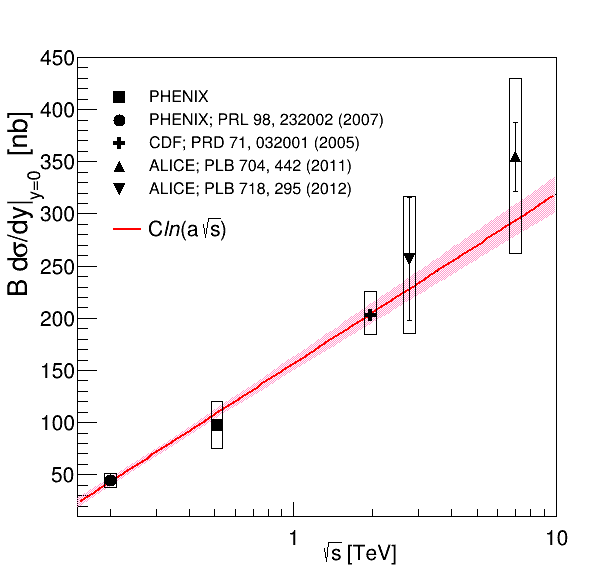}  
\caption{\label{fig:xsecworld}
The PHENIX results of $p_T$-integrated cross section times branching 
ratio for $J/\psi$ production at midrapidity, shown with world data. All 
systematic errors for CDF and ALICE experiments were added in 
quadrature. The fit parameters are C = 70.4 nb, and a = 9.27~TeV$^{-1}$.  
The pink band shows the one-sigma fit uncertainty.
}
\end{figure}

The $p_T$-integrated cross section times branching ratio is shown in 
Fig.~\ref{fig:xsecworld} along with the previous PHENIX results at 
$\sqrt{s}$ = 200~GeV~\cite{PHENIXcs} and the world results from the 
Large Hadron Collider~\cite{ALICEcs1,ALICEcs2} and 
Tevatron~\cite{CDFcs}. A simple logarithmic dependence on the collision 
energy is seen for $J/\psi$ production at midrapidity, making estimates 
of $J/\psi$ yield at any $\sqrt{s}$ easy, and inviting the theory 
community to model the trend. The $p_T$-integrated cross section times 
branching ratio was found to be 97.6 $\pm$ 3.6 (stat) $\pm$ 5.1 (syst) 
$\pm$ 9.8 (global) $\pm$ 19.5 (multiple collision) nb.

\begin{figure}[!ht]
\includegraphics[width=1.0\linewidth]{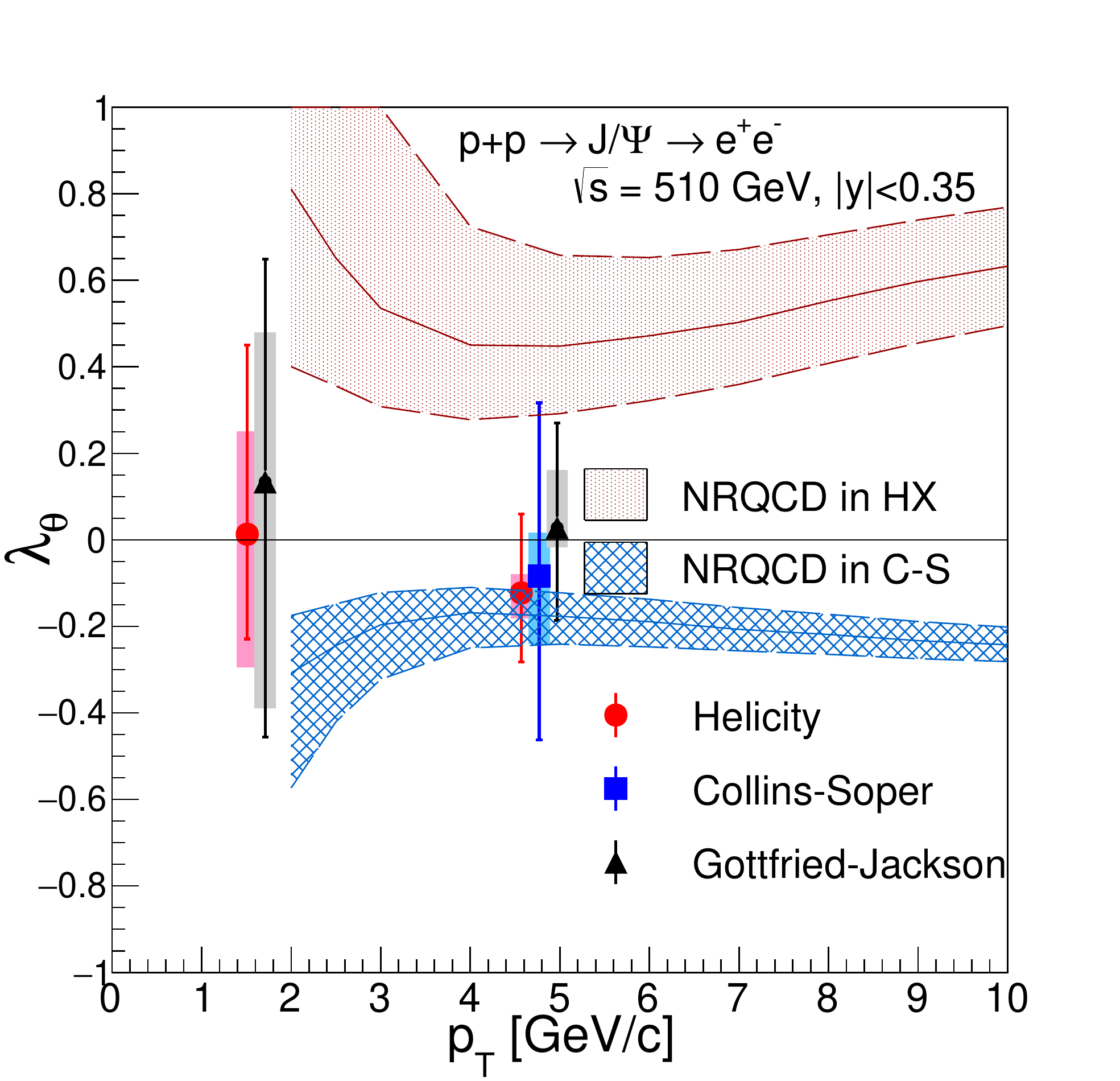}  
\caption{\label{fig:lamthe}
$\lambda_{\theta}$ measured in $J/\psi$ transverse momentum bins of 0.0 
$< p_T <$ 3.0~GeV/$c$ and 3.0 $\leq p_T <$ 10.0~GeV/$c$ overlaid with 
NRQCD predictions in the Helicity and Collins-Soper frames. The points 
for different frames are shifted for visual clarity.
}
\end{figure}

\begin{figure}[!ht]
\includegraphics[width=1.0\linewidth]{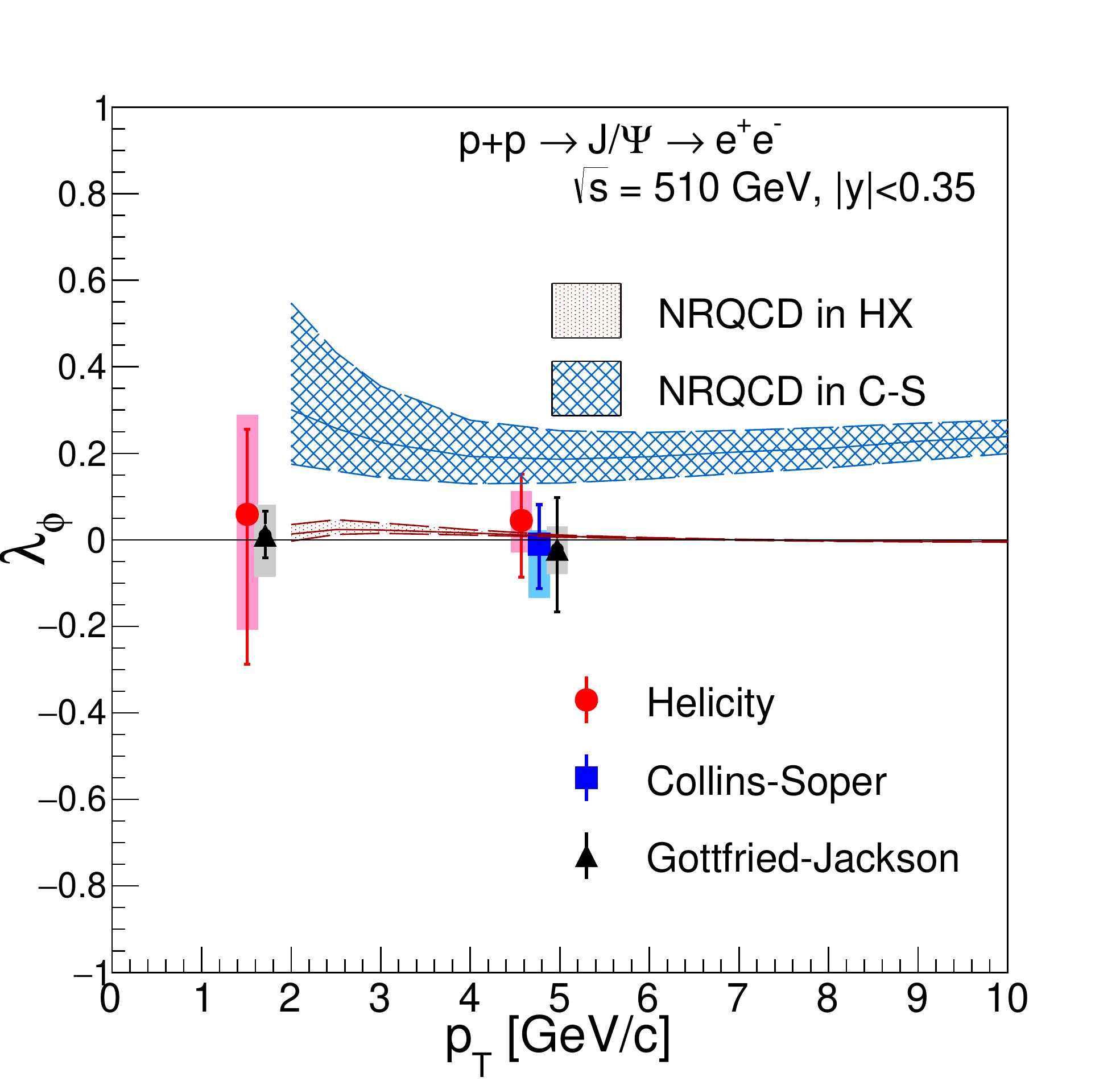}  
\caption{\label{fig:lamphi}
Angular coefficient $\lambda_{\phi}$ measured in $J/\psi$ transverse 
momentum bins of 0.0 $< p_T <$ 3.0~GeV/$c$ and 3.0 $\leq p_T <$ 
10.0~GeV/$c$ overlaid with NRQCD predictions in the Helicity and 
Collins-Soper frames. The points for different frames are shifted for 
visual clarity.
}
\end{figure}

\begin{figure}[!th]
\includegraphics[width=1.0\linewidth]{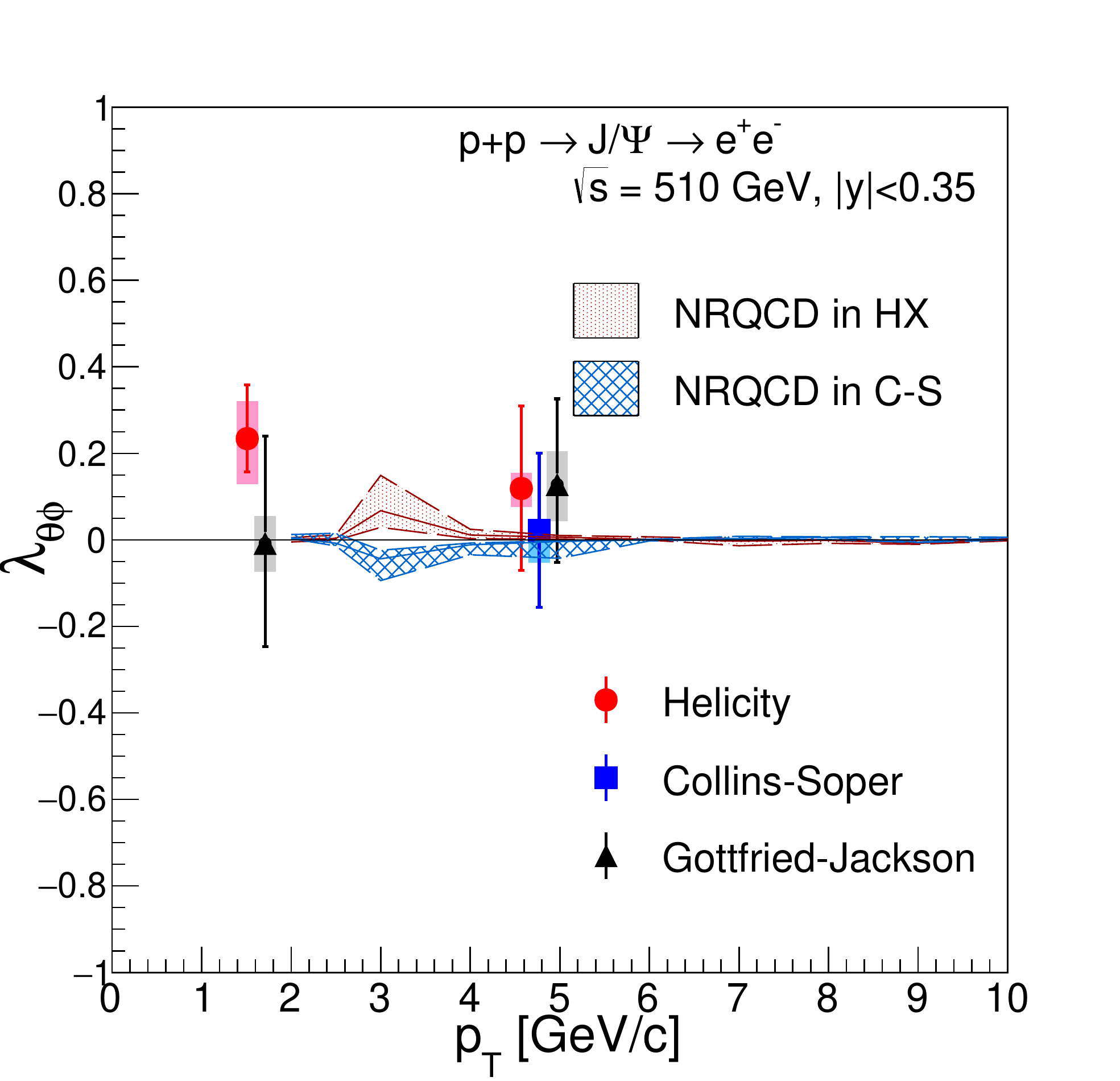}  
\caption{\label{fig:lamthephi}
Angular coefficient $\lambda_{\theta\phi}$ measured in $J/\psi$ 
transverse momentum bins of 0.0 $< p_T <$ 3.0~GeV/$c$ and 3.0 $\leq p_T 
<$ 10.0~GeV/$c$ overlaid with NRQCD predictions in the Helicity and 
Collins-Soper frames. The points for different frames are shifted for 
visual clarity.
}
\end{figure}

The final results of the three primary angular coefficients are shown in 
Fig.~\ref{fig:lamthe}, Fig.~\ref{fig:lamphi}, and 
Fig.~\ref{fig:lamthephi}. Uncertainties on measurements in different 
frames are correlated.  Due to limited detector acceptance in $\eta$ at 
midrapidity, $\lambda_{\theta}$ (Fig.~\ref{fig:lamthe}) is poorly 
constrained compared to $\lambda_{\phi}$ (Fig.~\ref{fig:lamphi}).  
Theory predictions based on full NRQCD at NLO with leading relativistic 
corrections that includes CS and CO states, provided by Butensch\"on et 
al.~\cite{Butenschoen:2012}, are overlaid with the measurements. The 
uncertainty bands on the NRQCD predictions account for the scale 
uncertainties and uncertainties on the LDMEs. The LDMEs were obtained in 
a global analysis of unpolarized data that excludes measurements from 
hadroproduction with $p_T<$3.0~GeV/$c$. To improve consistency with the 
data, the feed-down from heavier charmonium states was subtracted in the 
theory prediction. The fraction of $J/\psi$ events from $b$-flavored 
hadron decays is negligible at RHIC. In PHENIX, the unpolarized yield 
measurements~\cite{Adare:2010tj} are well described down to 1~GeV/$c$, 
justifying the comparison to the polarized measurements.

\begin{figure}[!thb]
\includegraphics[width=1.0\linewidth]{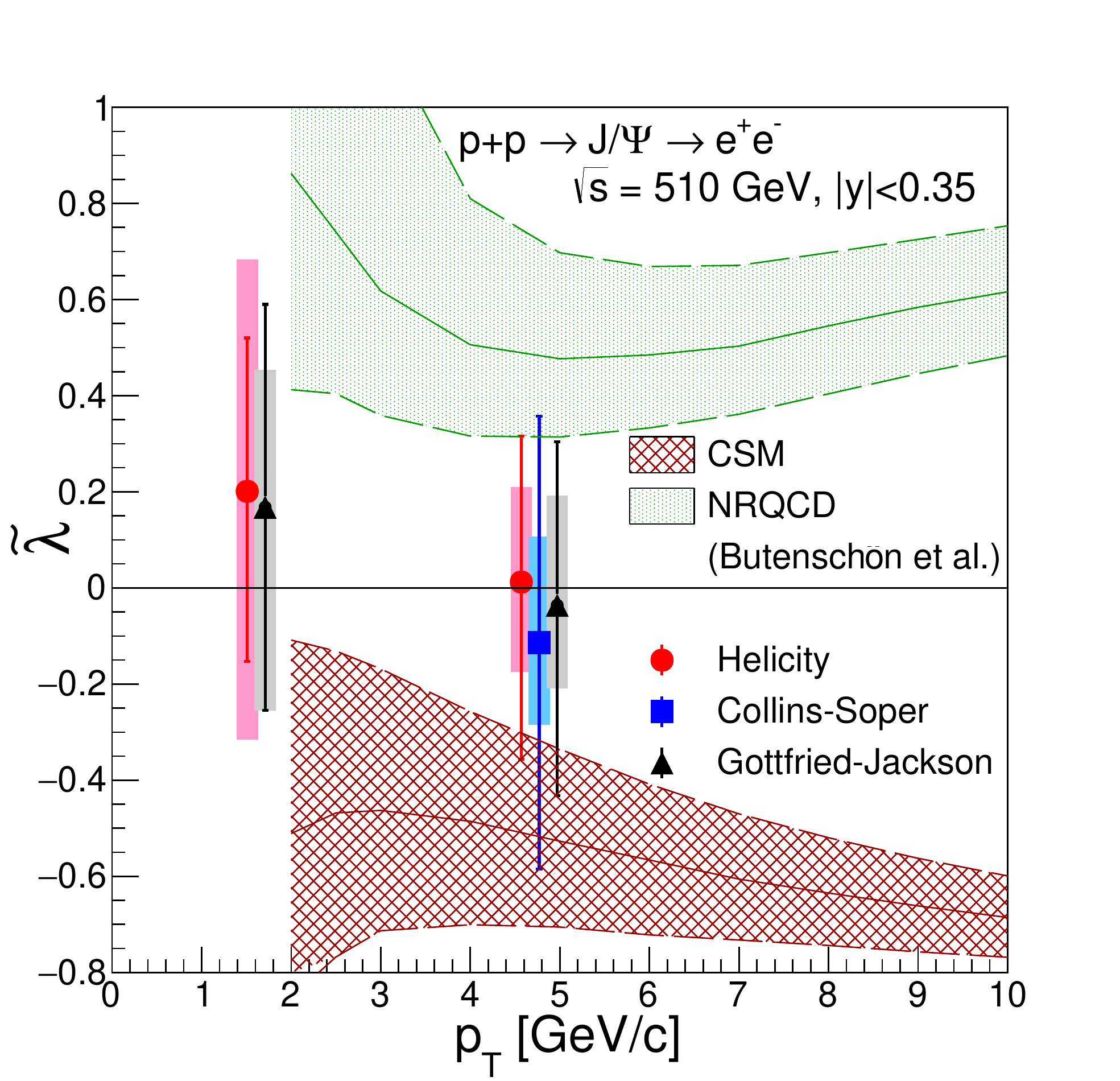}  
\caption{\label{fig:lamtil}
$\tilde{\lambda}$ measured in $J/\psi$ transverse momentum bins of 0.0 
$< p_T <$ 3.0~GeV/$c$ and 3.0 $\leq p_T <$ 10.0~GeV/$c$ overlaid with 
CSM and NRQCD predictions in the Helicity frame. Predictions for this 
frame-invariant variable in the other two frames are consistent with the 
one in the Helicity frame. The points for different frames are shifted 
for visual clarity.
}
\end{figure} 

\begin{figure}[!thb]
\includegraphics[width=1.0\linewidth]{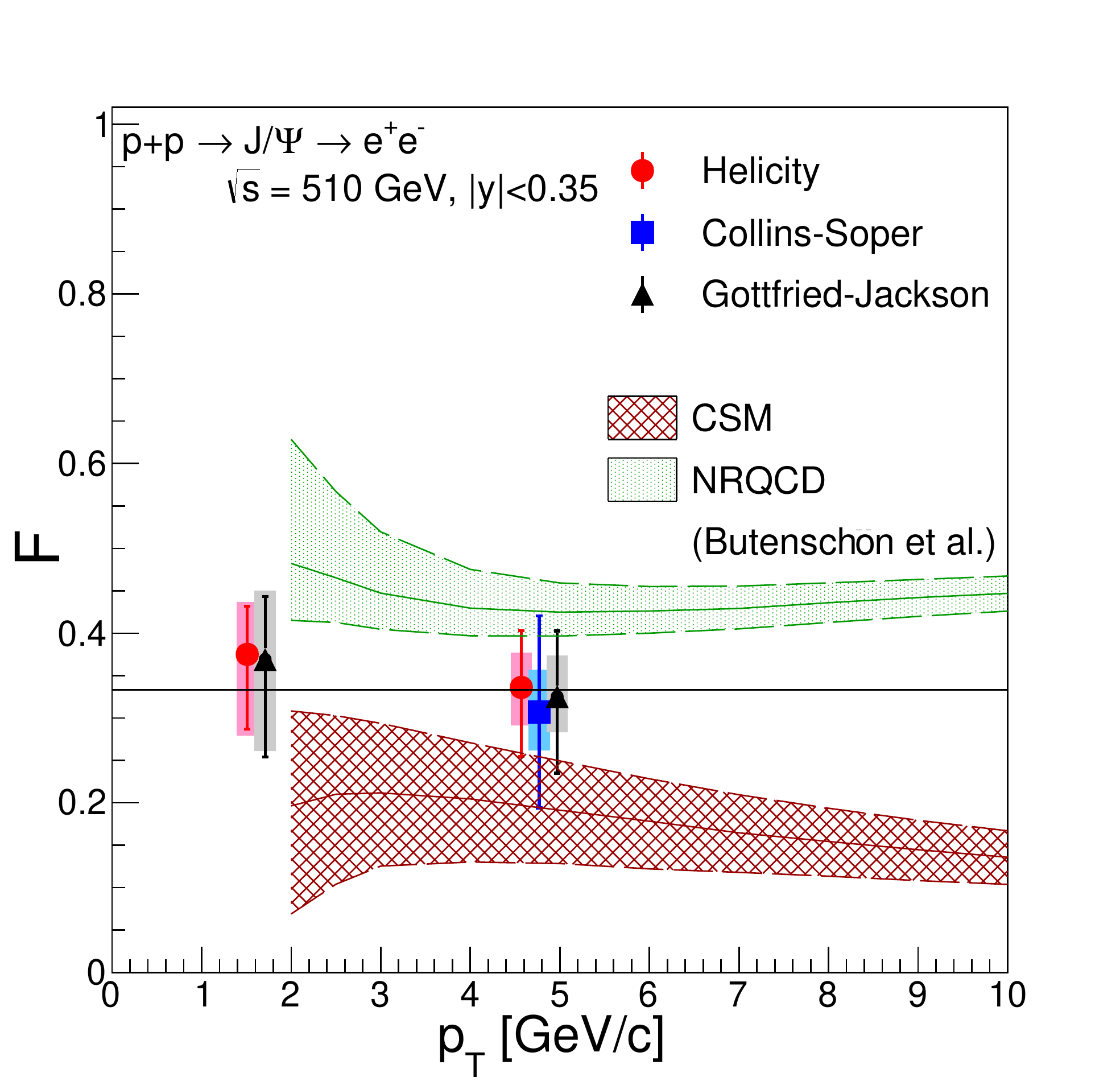}  
\caption{\label{fig:invf}
Invariant F measured in $J/\psi$ transverse momentum bins of 0.0 $< p_T 
<$ 3.0~GeV/$c$ and 3.0 $\leq p_T <$ 10.0~GeV/$c$ overlaid with CSM and 
NRQCD predictions in the Helicity frame. Predictions for this 
frame-invariant variable in the other two frames are consistent with the 
one in the Helicity frame. The points for different frames are shifted 
for visual clarity.
}
\end{figure} 

In Figs.~\ref{fig:lamtil} and \ref{fig:invf}, the results are also shown 
in terms of frame-invariant observables $\tilde{\lambda}$ and F, defined 
in Eq.~\ref{eq:rotinvvar}. Measuring these invariant variables provides 
a direct test for the underlying production mechanisms. Comparing 
results in different frames can additionally serve as robust tools to 
address systematic uncertainties. Consistency between different frames 
shown for $\tilde{\lambda}$ in Fig.~\ref{fig:lamtil} and for F in 
Fig.~\ref{fig:invf} indicates that systematic effects are under good 
control. Frame-invariant variables are interpreted as the total net 
polarization summed over all production mechanisms. Different production 
mechanisms can lead to natural polarization in different frames. In the 
case of $\tilde{\lambda}$, transverse polarization corresponds to the 
value of 1 and longitudinal polarization to the value of -1. The zero 
value seen in $\tilde{\lambda}$ is interpreted as no net polarization. 
The other variable F also carries similar meaning. Unlike 
$\tilde{\lambda}$, F is mathematically bounded between 0 and 1. The zero 
polarization corresponds to $\frac{1}{3}$ and transverse and 
longitudinal polarization to $\frac{2}{3}$ and 0, respectively.

Two scenarios were considered for these frame-invariant results, NRQCD 
and the Color Singlet Model, which is equivalent to the $v\rightarrow 0$ 
limit of NRQCD. The uncertainties on CSM predictions include the scale 
uncertainty. The CSM prediction is qualitatively opposite to the NRQCD 
predictions~\cite{Butenschoen:2012,Butenschoen:2013}. NRQCD predicts 
transverse polarization while the CSM predicts longitudinal 
polarization, as is shown in Fig.~\ref{fig:lamtil} and 
Fig.~\ref{fig:invf}. The data does not conclusively exclude either 
strong transverse or longitudinal polarization.

\begin{figure}[!thb]
\includegraphics[width=1.0\linewidth]{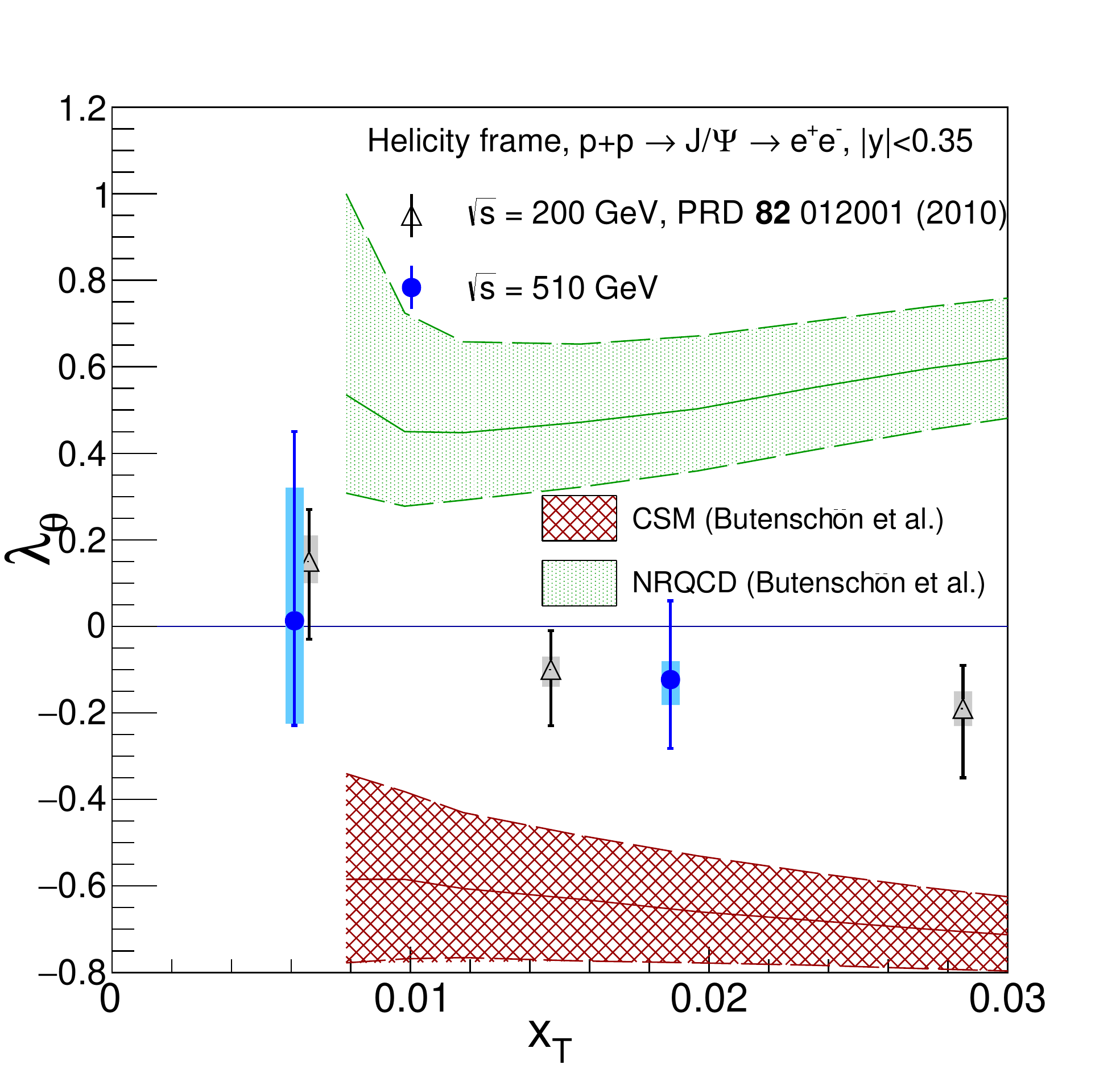}  
\caption{\label{fig:lamthemidrapcomp}
Angular coefficient $\lambda_{\theta}$ of midrapidity $J/\psi$ 
production at $\sqrt{s}$= 200~GeV and 510~GeV, shown as a function of 
$x_T = \frac{2 p_{T}}{\sqrt{s}}$. Data points at the lowest $p_T$ have 
been shifted against each other for visual clarity.
}
\end{figure}

\begin{figure*}[!thb]
\begin{minipage}{0.48\linewidth}
\includegraphics[width=0.99\linewidth]{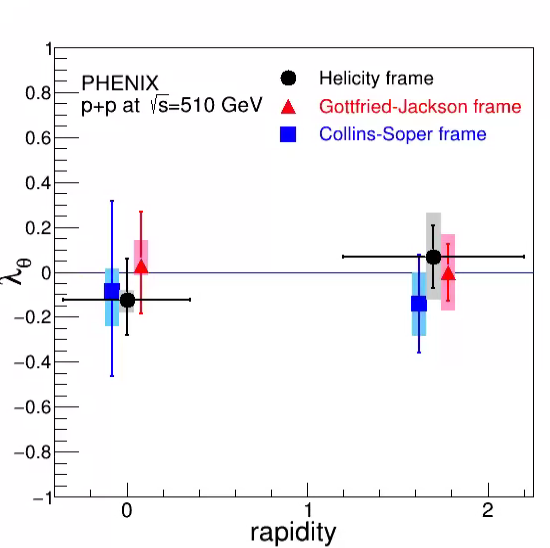}  
\caption{\label{fig:lamtherapcomp}
Angular coefficient $\lambda_{\theta}$ of $J/\psi$ production at 
$\sqrt{s}$ = 510~GeV shown as a function of rapidity in three 
polarization frames. The points for different frames are shifted for 
clarity. Forward rapidity points are from Ref.~\protect\cite{Adare:2016jta}.
}
\end{minipage}
\hspace{0.4cm}
\begin{minipage}{0.48\linewidth}
\includegraphics[width=0.99\linewidth]{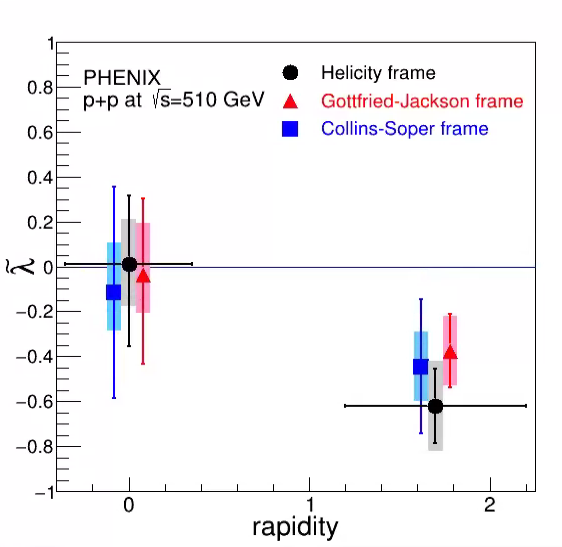}  
\caption{\label{fig:lamtilrapcomp}
Rapidity dependence of $\tilde{\lambda}$ of $J/\psi$ production at 
$\sqrt{s}$ = 510~GeV in three polarization frames. The points for 
different frames are shifted for clarity. Forward rapidity points are 
from Ref.~\protect\cite{Adare:2016jta}.
}
\end{minipage}
\end{figure*}

\begingroup
\begin{table*}[tbh]
\caption{\label{table:polresults}
Results with statistical $\oplus$ total systematic uncertainties.
}
\begin{ruledtabular}
\renewcommand{\arraystretch}{2}
\begin{tabular}{ccccccc}
Frame & $p_T$ & $\lambda_{\theta}$  &  $\lambda_{\phi}$  &
$\lambda_{\theta\phi}$  & $\tilde{\lambda}$ & F  \\
\hline
Helicity
& 0.0$<p_T<$3.0
& 0.01$^{+(0.44 \oplus 0.24) }_{-(0.24 \oplus 0.31)}$
& 0.06$^{+(0.20 \oplus 0.23)}_{ -(0.35\oplus 0.27)}$
& 0.23 $^{+(0.12 \oplus 0.09)}_{-(0.08\oplus 0.11)}$
& 0.20 $^{+(0.32 \oplus 0.48)}_{-(0.35 \oplus 0.52)}$
& 0.38 $^{+(0.06\oplus 0.06)}_ {-(0.09 \oplus 0.10)}$ 
\\
& 3.0$\leq p_T<$10.0
& -0.12$^{+(0.18\oplus 0.04)}_{-(0.16\oplus 0.06)}$
& 0.06$^{+(0.11\oplus 0.07)}_{-(0.13\oplus 0.07)}$
& 0.12$^{+(0.19\oplus 0.04)}_{-(0.19 \oplus 0.04)}$
&0.01$^{+(0.30\oplus 0.20 )}_{-(0.37 \oplus 0.19)}$
&0.34 $^{+(0.07\oplus 0.04)}_{-(0.08 \oplus 0.05)}$
\\
Collins-Soper
& 3.0$\leq p_T<$10.0
& -0.08$^{+(0.40\oplus 0.10)}_{-(0.38 \oplus 0.16)}$
& -0.01$^{+(0.09\oplus 0.03)}_{-(0.10 \oplus 0.12)}$
& 0.02$^{+(0.18 \oplus 0.01 )}_{-(0.18 \oplus 0.07)}$
& -0.11$^{+(0.47 \oplus 0.22)}_{-(0.47 \oplus 0.17)}$
& 0.31$^{+(0.11 \oplus 0.05 )}_{-(0.11 \oplus 0.05)}$
\\
Gottfried-Jackson
&  0.0$<p_T<$3.0
& 0.13 $^{+(0.52 \oplus 0.35)}_{-(0.59\oplus 0.52)}$
& 0.01 $^{+(0.06 \oplus 0.07)}_{-(0.05 \oplus 0.10)}$
& -0.01 $^{+(0.25\oplus 0.06)}_{-(0.24\oplus 0.07)}$
&  0.17$^{+(0.42\oplus 0.29)}_{-(0.42\oplus 0.42)}$
&0.37$^{+(0.07\oplus 0.08)  }_{-(0.12 \oplus 0.11)}$
\\
& 3.0$\leq p_T<$10.0
& 0.03$^{+(0.24 \oplus 0.13)}_{-(0.21\oplus 0.05) }$
& -0.02$^{+(0.12\oplus 0.05)}_{-(0.15 \oplus 0.06)}$
& 0.13$^{+(0.20\oplus 0.08)}_{-(0.18 \oplus 0.09)}$
& -0.04$^{+(0.34\oplus 0.23)}_{-(0.40\oplus 0.17)}$
& 0.33$^{+(0.08 \oplus 0.05)}_{-(0.09 \oplus 0.04)}$
\\
\end{tabular} \end{ruledtabular}
\end{table*}
\endgroup

Results are also compared to previous PHENIX measurements at a different 
beam energy~\cite{Adare:2010tj} and rapidity~\cite{Adare:2016jta}. 
Figure~\ref{fig:lamthemidrapcomp} compares $\lambda_{\theta}$ in the 
helicity frame for different collision energies. At midrapidity, the 
decay lepton spin alignment is consistent with no polarization both at 
$\sqrt{s}$ = 200~GeV (1-dimensional analysis) and 510~GeV (2-dimensional 
analysis). The rapidity dependence of $\lambda_{\theta}$ and 
$\tilde{\lambda}$ is shown in Fig.~\ref{fig:lamtherapcomp} and 
Fig.~\ref{fig:lamtilrapcomp}, respectively. While midrapidity data 
indicate no polarization, moderate polarization is seen at forward 
rapidity. At forward rapidity $\tilde{\lambda}$ was measured to be 
largely negative, indicating longitudinal polarization. This is in stark 
contrast to this result that sees no preferred polarization direction, 
shown in Fig.~\ref{fig:lamtherapcomp} and Fig.~\ref{fig:lamtilrapcomp}. 
No strong polarization was seen in other experiments at higher $p_T$ and 
higher beam energies in general, and the discrepancy between 
measurements and theory predictions is still being studied. Results of 
polarization measurements are summarized in 
Table~\ref{table:polresults}.

\section{Summary}

The PHENIX experiment measured the $J/\psi$ polarization at midrapidity 
in $\sqrt{s}=$ 510 GeV $p$$+$$p$ collisions by reconstructing the 
hadronized charmonium state in the dielectron decay channel. The 
midrapidity cross section at $\sqrt{s}=$ 510 GeV in the same channel has 
been newly measured and is consistent with NRQCD calculations above $p_T 
\gtrsim 2$~GeV/$c$. The results show the expected hardening of the 
$J/\psi$ $p_T$ spectrum as compared to the measurement at $\sqrt{s}=$ 
200 GeV.  At both low $p_T$ and high $p_T$, the net polarization has 
been observed to be consistent with zero within uncertainties. This is 
in contrast to the measurements made at forward rapidity. The new 
results do not rule out either the CSM or the NRQCD $J/\psi$ production 
models. The new measurements from the 2-dimensional analysis show 
consistency in $\lambda_\theta$ with the results from a previous 
1-dimensional midrapidity analysis at $\sqrt{s}$ = 200~GeV.



\begin{acknowledgments}

We thank the staff of the Collider-Accelerator and Physics
Departments at Brookhaven National Laboratory and the staff of
the other PHENIX participating institutions for their vital
contributions.  We acknowledge support from the
Office of Nuclear Physics in the
Office of Science of the Department of Energy,
the National Science Foundation,
Abilene Christian University Research Council,
Research Foundation of SUNY, and
Dean of the College of Arts and Sciences, Vanderbilt University
(U.S.A),
Ministry of Education, Culture, Sports, Science, and Technology
and the Japan Society for the Promotion of Science (Japan),
Conselho Nacional de Desenvolvimento Cient\'{\i}fico e
Tecnol{\'o}gico and Funda\c c{\~a}o de Amparo {\`a} Pesquisa do
Estado de S{\~a}o Paulo (Brazil),
Natural Science Foundation of China (People's Republic of China),
Croatian Science Foundation and
Ministry of Science and Education (Croatia),
Ministry of Education, Youth and Sports (Czech Republic),
Centre National de la Recherche Scientifique, Commissariat
{\`a} l'{\'E}nergie Atomique, and Institut National de Physique
Nucl{\'e}aire et de Physique des Particules (France),
Bundesministerium f\"ur Bildung und Forschung, Deutscher Akademischer
Austausch Dienst, and Alexander von Humboldt Stiftung (Germany),
J. Bolyai Research Scholarship, EFOP, the New National Excellence
Program ({\'U}NKP), NKFIH, and OTKA (Hungary),
Department of Atomic Energy and Department of Science and Technology
(India),
Israel Science Foundation (Israel),
Basic Science Research and SRC(CENuM) Programs through NRF
funded by the Ministry of Education and the Ministry of
Science and ICT (Korea).
Physics Department, Lahore University of Management Sciences (Pakistan),
Ministry of Education and Science, Russian Academy of Sciences,
Federal Agency of Atomic Energy (Russia),
VR and Wallenberg Foundation (Sweden),
the U.S. Civilian Research and Development Foundation for the
Independent States of the Former Soviet Union,
the Hungarian American Enterprise Scholarship Fund,
the US-Hungarian Fulbright Foundation,
and the US-Israel Binational Science Foundation.

\end{acknowledgments}




%
 
\end{document}